\documentstyle{sup2}
\include{psfig}
\vspace{-8pt}

\title[Superlattices and Microstructures, Vol.\ ??, No.\ ??, 1999]
{ Controlling Josephson transport by manipulation of Andreev levels in ballistic mesoscopic junctions }

\author[Superlattices and Microstructures, Vol.\ ??, No.\ ??, 1999]
{G\"{O}RAN WENDIN, VITALY S. SHUMEIKO, PETER SAMUELSSON \cr\vspace{8pt}
{\normalsize\it Chalmers University of Technology and G\"oteborg University,
S-412 96 G\"oteborg, Sweden}\cr
}

\begin{document}
\label{firstpage}
\maketitle
\sloppy
\begin{center}
\received{(Submitted 6 December 1998)}
\end{center}

\begin{abstract}
We discuss how to control dc Josephson current by influencing the structure and nonequilibrium population of Andreev levels via external electrostatic gates, current injection and electromagnetic radiation. In particular we will consider the "giant" Josephson critical current in "long" SIS tunnel junctions and the regular and anomalous nonequilibrium Josepson currents in three terminal SNS junctions. We will briefly discuss applications to the Josephson field effect transistor (JOFET) and to the newly invented Josephson interference transistor (JOINT).
\end{abstract}

\section{Introduction}

The ability to control Josephson current in SNS and SIS junctions is certainly of fundamental interest, and may also become important in applications to superconducting electronics with multiterminal devices - superconducting transistors \cite{vanHout91,TakaAPL96,MKW98,SMNLHCHL98,BMWK98}.
The fundamental mechanism involves coherent transport of quasiparticles through the normal region between superconducting electrodes, influencing the more or less complex normal region via electrostatic gates, current injection and electromagnetic radiation. From an applied point of view, the goal is to switch the Josephson current with sufficient gain to be able to drive networks of components.
In recent experiments on diffusive \cite{MKW98} and ballistic
\cite{SMNLHCHL98} short multiterminal junctions, the Josephson current has been supressed by means of electron injection. All this makes it highly interesting to get a complete picture of nonequilibrium current transport in ballistic superconducting junctions.

In the following we will use the more general terminology of SXS to characterize normal regions X with energy dependent transmission amplitudes for electrons and holes, e.g. describing situations intermediate between SNS and SIS, or entirely different situations involving resonant double barrier structures (Fig. 1). 

\begin{figure}
\begin{centering}
\psfig{figure=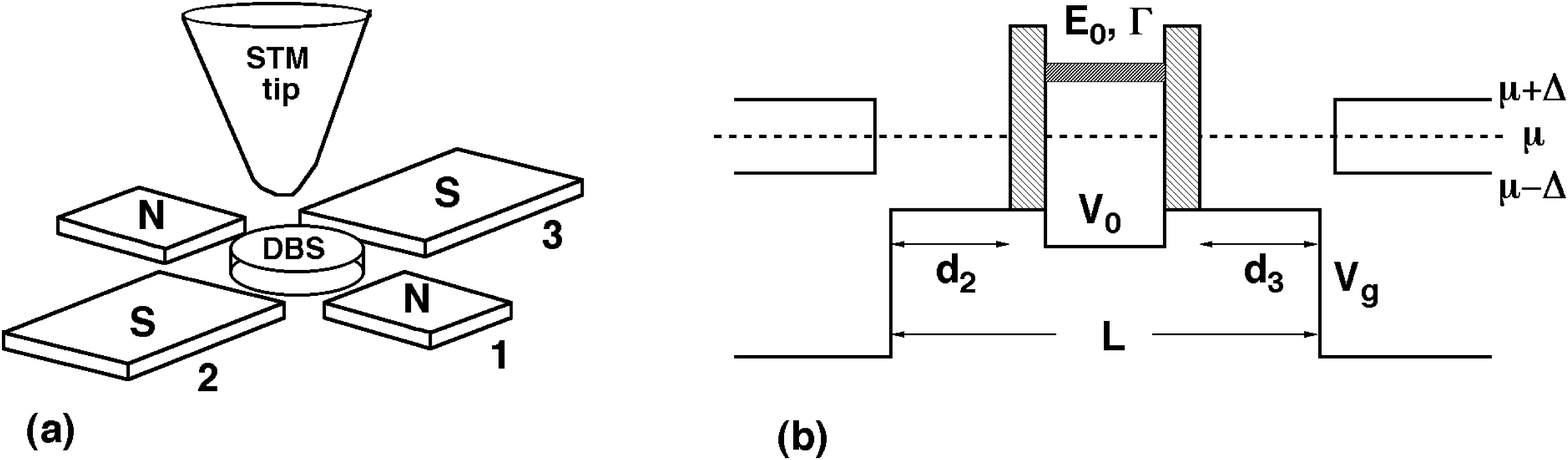,height=4cm}
\caption{(a) Schematic picture of a general complex SXS junction with many control possibilities. (b) 1D model of a "gate-controlled" SN-DBS-NS junction. If N=2DEG (2D electron gas), the broad potential barrier $V_g$ (length L) will include an effective barrier from the S-2DEG wave vector and effective mass mismatch. $V_0$ represents a control gate voltage for tuning the resonance properties of the double barrier structure (DBS).}
\end{centering}
\end{figure}

Within the framework of the 1D Bogoliubov-de Gennes equation a general dispersion equation for the bound Andreev level spectrum $E=E(\phi)$ as a function of the superconducting phase difference $\phi$ can be derived in terms of the normal energy dependent scattering amplitudes of the junction \cite{WeShu96}. Knowing the Andreev level spectrum, the Josephson current flowing through the junction through these bound states is then given by 
\begin{equation}
I=\sum_{bound} I(E)n(E), \;\;I(E)={2e\over\hbar}(dE/d\phi),
\end{equation}
where $n(E)$ is the Andreev level filling factor \cite{SBW97}.

Control of dc Josephson current in SXS junctions can be achieved either (i) by modifying the Andreev level structure by electrostatic gates, or (ii) by modifying the quasiparticle distribution and Andreev levels via injection of current or electromagnetic (EM) radiation into the normal region.
In order to achieve strong effects and efficient control, it is essential to taylor the superconducting junctions to have Andreev level spectra which are sensitive to external influence. One natural idea is then to utilize various resonance effects.

In this paper we will emphasize two types of coherent resonant transport through Andreev levels in ballistic junctions: (i) coupling of degenerate SN-interface Andreev levels through a tunnel barrier, and (ii) transport through a normal resonant level in a double barrier structure (DBS). 

Let us first consider an intermediate junction on the scale of the coherence length, with a barrier long enough to be completely non-transparent when it exceeds the Fermi level.
In this case, the gate-controlled long potential barrier  \cite{barrier} is always below the Fermi level, influencing the wave functions over a wide region with many $k_F$-oscillations, $k_FL\gg1$, with "free" electron-hole propagation and dephasing in the normal X-region under the gate.

When the gate potential $V_g$ increases towards the Fermi level, the Fermi  wave vector $k_F$, the electron density and the coherence length $\xi_N$ decrease. Simultaneously, the electron-hole dephasing increases, and the junction therefore effectively becomes long. This is reflected in the Andreev level structure, the gap successively filling up with Andreev levels, similar to the case of actually increasing the physical length $L$ of the SNS junction.

If this barrier is pushed above the Fermi level the transparency will vanish, $D \rightarrow 0$, and the junction will essentially become divided into two isolated SI systems, each having a localized Andreev surface/interface state sitting in the gap close to each gap edge \cite{WeShu95}. If we now decrease the length of the barrier to allow tunneling, $D \ll 1$ but finite, the degenerate SI interface states begin to interact through the tunnel barrier, and we get an SIS junction.

This then takes us to the next case: an SXS junction with  $N<X<I$, using a gate-controlled "tunnel" barrier to tune the junction from SNS to SIS. As long as the top of the low-transparency barrier does not pass above the Fermi level $\mu$, the junction will essentially behave as a short SNS junction, with a symmetric pair of dispersing Andreev levels around the Fermi level. Not until the barrier becomes significantly higher than the Fermi level does the junction approach SIS behaviour \cite{ WeShu95, Furusaki92}.

If the barrrier is long enough, for $D \rightarrow 0$ it can split the Andreev envelope into two degenerate parts localized around the respective SI (or SNI) interfaces - Andreev interface/surface states - a double-well situation. Weak coupling splits the degenerate Andreev levels into two bonding-antibonding levels, with level splitting $\sim \sqrt{D}$ - corresponding to resonance coupling of the superconducting electrodes.
The central aspect is that even in SIS junctions the development of the Andreev level structure depends not only on the strength of the barrier (via the transmission probability D), but also on the length of the barrier via electron-hole dephasing.

A second method of controlling dc Josephson current is connected with creating nonequilibrium population of Andreev levels. One way to achive this is by connecting the normal region to a normal reservoir with a shifted chemical potential (terminal 1 in Fig. 1a) \cite{Vanwees91, Volkov95, ChangBag97, SSW97, IlBag98, WSZ98, SSW99}. This opens up the junction, broadens the Andreev levels and modifies the wave functions of the Andreev states. Seen from the normal reservoir, the junction looks like a split NS Andreev mirror with resonances. This creates an anomalous interference current \cite{SSW97} which in long junctions is independent of junction length, scales linearly with voltage $V$ up to $\Delta$ and saturates at a magnitude that is of the order of the short junction current $I\sim e\Delta/\hbar$.

Another way to achieve nonequilibrium population of Andreev levels is via optical pumping using EM radiation.
Resonant pumping between two Andreev levels will strongly mix the wave functions
of these two levels, inducing highly coherent combinations of static Andreev states with pronounced interference effects, leading to Rabi oscillation of the Josephson current \cite{SWB93,Gor95}. This makes it possible selectively to control the population $n(E)$ of the resonant levels. By saturating (equalizing) the level populations one may e.g. induce suppression of the contribution of the resonant levels to the total current.

\section{Gate-potential control of Andreev levels - JOFET}

Takyayanagi and coworkers \cite{TakaAPL96} have recently demonstrated that the dc Josephson current in gate-controlled ballistic S-2DEG-S junctions can be suppressed using a gate voltage of only $V_g=-1V$, the entire current-voltage characteristics (IVC) transforming from RSJ to ohmic behaviour.

To describe this situation, consider a gate-controlled SNS-type junction with length $L \approx \xi_N$. This can be illustrated by Fig. 1b with the DBS removed, denoting the remaining flat potential barrier by $V_g$.
Since the potential barrier is broad, $k_FL\gg1$, it influences the wave functions over a wide region, defining a local Fermi wave vector and a coherence length for the normal region. 

To begin with, we set the effective mass $m^*=1$ everywhere, and the Fermi velocities are taken to be equal when $V_g=0$ (perfectly matched SNS). The  parameters are chosen so that the pure SNS junction ($V_g=0$) is fairly close to the short limit ($\xi_N \approx L$), with 1-2 bound Andreev levels localized in the N-region.

The coherence length in the normal region is defined as
$\xi_N^{-1} = \delta k = (k^+-k^-)$ where
$k^\pm=\sqrt{2m^*(\mu-V_g\pm E)}$. In the present case of fixed length L, the electron-hole dephasing is directly controlled by the Fermi vave vector in the gate region. For $E \ll \mu -V_g$, the
dephasing $\beta = \delta k L$ in the normal region is given by
\begin{equation}
\beta = EL/\sqrt{2m^*(\mu-V_g)}\; ; \;\; 0 < E \le \sqrt{\Delta}
\end{equation}
When the gate potential increases towards the Fermi level the electron-hole phase difference keeps increasing, allowing more and more Andreev levels to enter the gap from the continuum. This is precisely also what happens if one increases the length of the junction $L$ while keeping the potential fixed. Therefore, increasing the gate potential decreases the coherence length and effectiely makes the junction long.

We find that for increasing gate voltage $V_g$, the junction remains fairly short (2-3 Andreev levels) until $V_g/\mu \sim 0.9$, the critical current $I_c$ and the  $I_cR_N$ product decreasing slowly. 
However, for increasing gate voltage in the range $0.9 \le V_g/\mu \le 1-\Delta/\mu$, the gap region becomes raidly filled up by an increasing number of Andreev levels, causing the critical current $I_c$ and the $I_cR_N$ product rapidly to drop, the coherence length $\xi_N$ becoming short and the junction effectively becoming long, similar to the dirty case \cite {VolTakPRB96}.

When $1-\Delta/\mu \le V_g/\mu \le 1$, the top of the barrier enters the bottom of the superconducting gap $\Delta$ and approches the Fermi level, gradually depleting
the Andreev levels in the gap (symmetrically around the Fermi level), making the critical current $I_c$ and the $I_cR_N$ product approach zero.

Finally, when $V_g/\mu \ge 1$, the barrier passes the Fermi level,  the junction goes into the SIS tunnel regime and the critical current $I_c$ goes exponentially to zero. Since the barrrier is very long, in practice there is an  abrupt cutoff. Here is a difference in principle with the physically long junction where the gap keeps filling up, the bound-state current decaying like $\xi_N/L$.

In numerical studies of S-Semiconductor-S junctions \cite{Chrestin94,QDS98}, appropriate for the S-2DEG-S JOFET \cite{TakaAPL96}, the small effective mass ($m^*=0.04$) in the 2DEG normal region introduces a big potential barrier already in the absense of gate voltage. As a result there are pronounced normal resonances in the X-region which strongly influence the Andreev level spectrum and the Josephson current \cite {Chrestin94}, as described in Sect. 3. In particular, the $I_cR_N$ product is reduced already for zero gate voltage, $V_g=0$, in qualitative agreement with experiment \cite {QDS98}.

\section{Resonance Josephson coupling in complex junctions}

A general question we would like to address is: How can the Andreev level spectrum in an  S-X($V_g$)-S junction be influenced by a more or less complex barrier structure created by a split-gate arrangement, such as illustrated in Fig. 1?

The different parts of the potential barrier structure (Fig. 1b) are imagined to be controlled by by a split-gate combination such that one can independently vary  the broad potential $V_g$ (length L) and the position of the resonance in double barrier structure (DBS) via $V_0$. While we are at it, we could just as well imagine to be able also to control the barrier heights of the double barrier structure. As a result we can, in principle, control both the position and the width of the normal resonant level in the DBS.

Let us limit our discussion to short junctions. If we assume the narrow normal resonance to be far away from the Fermi level, $E_0\gg\Delta, \Gamma$, the DBS acts a low-transmission tunnel barrier. If the structure is symmetric, the Andreev level energies are given by ($|\phi-2\pi n|\gg \sqrt{\Delta/\mu}$)
\begin{equation}
E=\pm\Delta \sqrt{1-\left({\beta\over 2}\pm \sqrt{D}\sin{\phi\over 2}\right)^2},
\end{equation}
where $\beta = \beta(E) = \beta_{DBS} + \beta_2 + \beta_3$ is the total electron-hole dephasing due to transmission through the entire normal X-region (the double barrier structure (DBS) plus the two semiconducting or metallic regions; $\beta_i= 2d_i/\xi_N, i=2,3$). 

It follows from Eq. (3) that the current through individual Andreev levels is given by
\begin{equation}
I(E_\pm)= -sign(E_\pm) {e\Delta \over 2\hbar}  (D \sin\phi 
\pm {\beta\sqrt{D}\over 2}\cos{\phi\over2})
\end{equation}
If $\beta/2 \gg \sqrt{D}$, the second term in Eq. (4), originating from two Andreev bands with anomalously large dispersion, $E(\phi) \sim \sqrt{D}$, will dominate the current \cite{WeShu96, WeShu95}. Because of this large dispersion, the currents of the individual Andreev bound states will exceed the Ambegaokar-Baratoff critical current \cite{AB63}, the sum current in Eq. (4), by a factor $(\beta/2)/\sqrt{D}\gg1$. 

If there are no normal metal or semiconductor regions ($d_2=d_3=0$), the junction behaves like a SIS junction with transparency $D \ll 1$. For simplicity and rigour, let us describe this situation by a clean tunnel barrier \cite{WeShu95}. The dephasing parameter $\beta$ can then be directly related to the  electron-hole wave vector difference $\delta\kappa=\kappa^+ -\kappa^-$,  where $\kappa^\pm=\sqrt{2m^*(V_g-\mu \pm E)}$. To illustrate a typical situation, let us assume that the tunnel barrier is about twice as high as the Fermi energy, $V_g \approx 2\mu$. The dephasing parameter $\beta$ is then given by \cite{WeShu95}
\begin{equation}
\beta  \approx {\delta\kappa \over k_F} \ll 1
\end{equation}
which finally leads to the very simple condition for Andreev level splitting:
\begin{equation}
\sqrt{D} \ll {\delta\kappa \over k_F} \approx {\Delta \over \mu}  \ll 1.
\end{equation}

Let us emphasize that the form of the spectrum in Eq. (3) results from the coupling of degenerate Andreev surface/interface states \cite{WeShu96,WeShu95}.
If the DBS in Fig. 1b is made completely opaque, the junction will split into two isolated SNI/INS interfaces. For $D=0$, the energies of the degenerate Andreev SNI surface levels are given by $E=\pm\Delta \sqrt{1-\left(\beta/2\right)^2}$, where $\beta \approx \Delta/\mu + 2d/\xi_N$. 
Consider first the extreme SIS case: then $d=0$ and $\beta \approx \Delta/\mu \ll 1$, and the Andreev surface levels will be sitting very close to the gap edges. Increasing the width $d$, dephasing will be dominated by the normal interface regions: in this case $\beta=2d/\xi_N$ can be large, and the Andreev surface state will be pulled down into the gap.

Bringing together the degenerate SI surface states causes resonance tunneling through the barrier and the formation of a two-level system of "bonding-antibonding" Andreev levels split by  $\Delta \beta \sqrt{D}$, the Josephson resonance coupling.
The enhanced currents in Eq. (4) will appear as nonequilibrium currents in the junction: in equilibrium they will be hidden due to mutual cancellations.

Bringing the SI/IS interfaces closer together will finally increase the transparency $D$ so much that the Josephson coupling will dominate over electron-hole dephasing: the upper level becomes pushed into the continuum and the Andreev level begins to disperse on the scale of $\Delta D$: there will then only be a single level, $E=\pm\Delta\sqrt{1-D\sin^2\phi/2 }$ \cite{Furusaki92}.

Asymmetry in position of the Andreev surface levels  due to asymmetry of the junction geometry, $\delta \beta = \beta_1 -\beta_2 = 2 |d_1-d_2|/\xi_N > \sqrt{D}$, leads to a modificaiton of the Andreev level energy,
\begin{equation}
E=\pm\Delta \sqrt{1-\left({\beta\over 2} 
\pm \sqrt{(\delta\beta)^2+D\sin^2{\phi\over2}}\right)^2},
\end{equation}
This type of asymmetry, or asymmetry due to fluctuations 
of the superconducting order parameter
in the electrodes, $\Delta_1-\Delta_2 >\beta\sqrt{D}\Delta$, will 
destroy the resonance coupling and give back the usual expression $E=\pm\Delta\sqrt{1-D\sin^2\phi/2 }$.

The discussion above concerned a narrow level far away from the Fermi level, $E_0 \gg \Gamma$, describing non-reonant tunneling through the DBS structure. In the case of a narrow resonance near the Fermi level, $E_0 \le \Gamma$, we instead obtain \cite{WeShu96, BeenHout91}
\begin{equation}
E=\pm\sqrt{E_0^2+\Gamma^2\cos^2(\phi/2)}.
\end{equation}
When the resonance approaches the Fermi level, $E_0=0$, and the structure
becomes completely transparent for normal electrons. The Andreev energy and Josephson current 
\begin{equation}
E=\pm \Gamma\cos(\phi/2)\;;\;\; I=e\Gamma\sin(\phi/2)
\end{equation}
then have the form typical for an SNS junction, but with $\Delta$ replaced by $\Gamma$.

In a wave function picture, in the SNINS structure the Andreev bound state wave function is localized in the two N-regions, and hybridization through the I-barrier (non-resonant DBS) causes resonant hopping of quasiparticles and determines the bandwidth (Andreev level splitting). When a normal Breit-Wigner resonance in the DBS region is pulled towards the Fermi level, the Andreev bound state wave function moves into the resonant central N-region: resonance hopping of quasiparticles is then suppressed, and the bandwidth is determined by the width of the normal resonance.

Finally, it should be noted that d-wave superconductors with a nodeline along the SI surface normal are characterized by midgap (zero energy) Andreev surface states \cite{Hu93}. In  a SIS (or SNINS) junction, these Andreev states will interact, leading to resonance splitting $\sim \sqrt{D}$ and resonant transport \cite{RiedBag98,Wendin98}, as discussed above.

\section {Non-equilibrium population of Andreev levels via current injection - JOINT }

So far we have only considered equilibrium Josephson current in the 1D ballistic SXS junction in Fig. 1, because the injection lead 1 has been unbiased and essentially disconnected.

We now connect the injection lead 1 to a normal reservoir and apply a bias voltage $V_1=V$, thus broadening the Andreev levels and creating a nonequilibrium quasiparticle distribution.
The assumption is that the normal X-region of the junction will be in equilibrium with the normal reservoir, and therefore in {\em nonequilibrium relative to the superconducting electrodes}.

When the junction is open, the Andreev levels not only broaden but also split into two quasibound states associated with injected electrons and holes \cite{SSW97, SSW99}. The total nonequilibrium current in the junction is given by 
\begin{equation}
I=\int dE \left[i^e n^e +i^h n^h +i^s n \right]
\end{equation}
where $i^s$ is the quasiparticle current injected from the superconducting electrodes, and $i^e$, $i^h$ are currents of electrons and holes injected from the normal reservoir. These currents are controlled by the reservoir, $n^{e,h}=n(E\pm eV)$.
To maintain this control, the coupling $\varepsilon$
between the junction and the reservoir must be sufficiently strong: the
lifetime within the junction $d/(\varepsilon v_F)$ must be smaller than
inelastic relaxation time $\tau_i=l_i/v_F$, i.e. $\varepsilon>d/l_i$.

If we introduce sum and difference currents $i^\pm = i^e \pm i^h$,
the total current may be written as
\begin{equation}
I=I_{eq}+ I_{neq} = I_{eq}+I_r+I_a
\end{equation}
in terms of the equilibrium current 
$I_{eq}=\int (i^+ + i^s) n \;dE $, 
the regular nonequilibrium current 
$I_{r}=\int (i^+/2)(n^e+n^h-2n) \;dE$ 
and the anomalous nonequilibrium current 
$I_{a}=\int (i^-/2)(n^e-n^h) \;dE $.

In the nonequilibrium current $I_{neq}$, the regular term $I_r$ describes the effects of nonequilibrium population of the the Andreev levels, while the anomalous term $I_a$ describes modification of the Andreev states themselves due to the open injection lead. This modification is not primarily the broadening of the levels: much more important are the changes in the transition amplitudes for electrons and holes, leading to a difference current $i^-$ which alwas has the same sign, as seen in Fig. 2a.

If we assume the normal reservoir to be weakly coupled to the SNS-junction via a tunnel barrier, with  $\varepsilon$ a small coupling constant, in the limit $\varepsilon \ll 0$ the quasibound states develop into true bound states, Andreev levels, responsible for the current transport in the junction. In this limit we obtain the sum current 
\begin{equation}
i^+=\frac{2e}{\hbar}\frac{dE}{d\phi}\delta(E-E_n^{\pm}),
\end{equation}
and the difference current
\begin{equation}
i^-=
-{e\over 2\hbar} {\mbox{Im}(rd^\ast)\sin\phi \over \sqrt D |\cos(\phi/2)| 
\sqrt{1-D\sin^2(\phi/2)}} \left| {dE\over d\phi} \right| \delta(E-E_n^{\pm}).
\end{equation}
We emphasize that the difference current arises due to essential modification of the interference effects building up the Andreev states.

In the weak coupling limit $\varepsilon \ll 0$, the Andreev resonances become very narrow, and we may use the bound state expression $dE/d\phi$ to calculate all currents. An important observation is that the anomalous current is proportional to $|dE/d\phi|$, i.e the {\em modulus} of the
current carried by each level, as shown numerically in Fig. 2a. This means that the absolute values of all currents add up, preventing the cancellation between consecutive Andreev levels responsible for the exponentially small equilibrium current in long junctions.

\begin{figure}
\begin{centering}
\psfig{figure=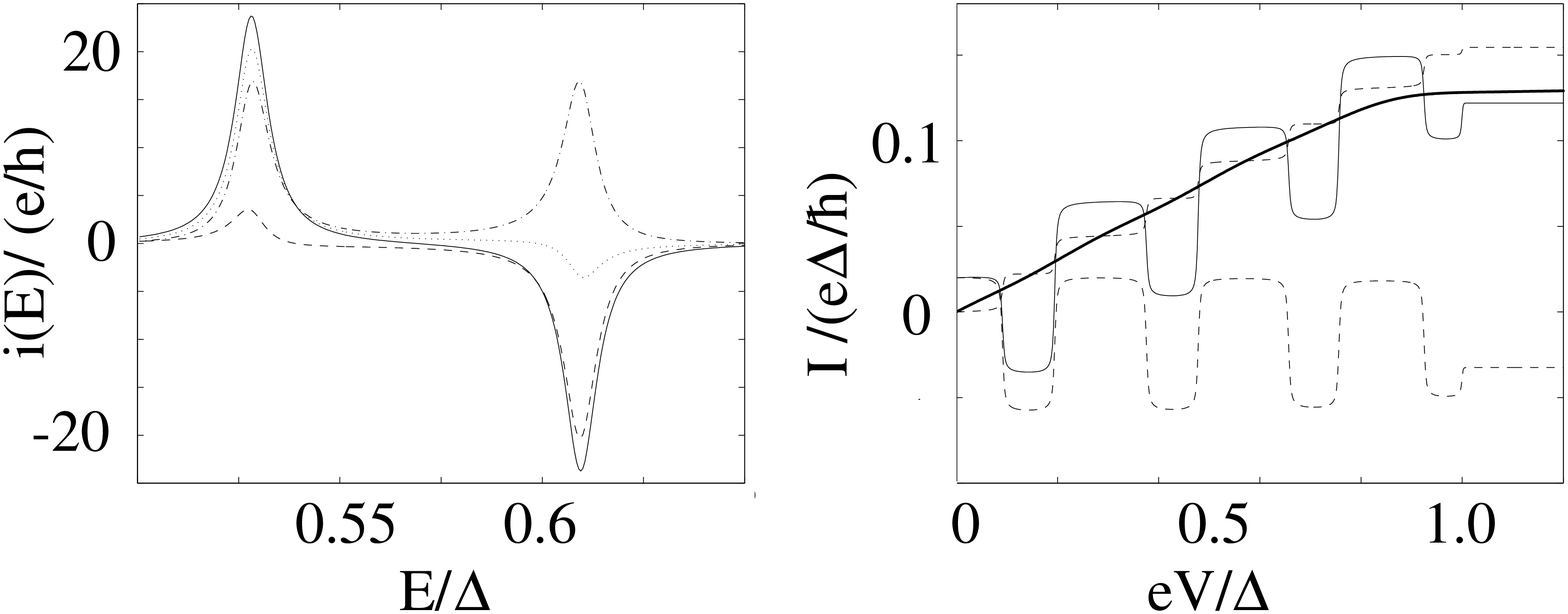,height=5cm}
\caption {
(a) The charge current densities $i^e$ (dotted), $i^h$
(dashed), $i^+$ (solid) and $i^-$ (dash-dotted) as functions of energy.
$L=20 \xi_0$, $D=0.8$, $\epsilon=0.1$ and $\phi=0.4 \pi$. 
(b) The currents $I_{eq}+I_r$, regular current (lower dasded), $I_a$, anomalous
current (upper dashed), and the sum $ I_{eq}+I_r+I_a$ (thin solid) at $T=0$.
$ I_{eq}+I_r+I_a$ at $kT=0.1 \Delta$  (thick solid). $L=20 \xi_0$, $D=0.8$,
$\epsilon=0.05$ and $\phi=0.4 \pi$.
}
\end{centering}
\end{figure}

The anomalous current is proportional to Im$(rd^*)$, i.e. is sensitive to the phase of the midpoint scatterer. The anomalous current therefore has interference origin and can be attributed to modification of the bound Andreev states themselves (i.e., not only to nonequilibrium population, responsible for the regular current) \cite{SSW97,SSW99}. The effect arises because the midpoint scatterer acts as a beam splitter at the injection point, creating left and right moving beams of inected electrons and holes. The current through the SXS junction then becomes a coherent superposition of split beams of electron and holes having scattered at the two SN interfaces, picking up superconducting phases $\pm\phi/2$ at the SN interfaces and transmission and reflection phase shifts at the midpoint scatterer (and at the SN interfaces if the normal interface transmission is not perfect). As a result, the difference current keeps the same direction for each Andreev level (for fixed phase of the scatterer).

Figure 2b shows the IVCs of the regular, anomalous and total nonequilibrium currents as as functions of bias voltage of the injection lead. Of particular interest is that the anomalous
current in long junctions at finite temperatures scales linearly with
voltage and saturates at $eV/\Delta$ at a current of the same magnitude as the short junction current. The equilibrium current of a short junction $L<<\xi_0$ can be supressed by the nonequilibrium regular current for $eV>\Delta$ (Fig. 2b).
For a long junction $L>>\xi_0$ at zero temperature, the equilibrium current oscillates for $eV<\Delta$, and is suppressed for $eV>\Delta$. The nonequilibrium current has a staircase-like voltage dependence and saturates at $eV>\Delta$ at a level typical for the current of a short junction. At finite temperature the equilibrium current is exponentially small, and the nonequilibrium current scales linearly with voltage for $eV<\Delta$.

Note that the injection current is proportional to the anomalous current, and scales with the coupling constant: $I_1 \sim \varepsilon \; I_a$. The anomalous current can therefore in principle be controlled with a very small injection current. The injection current is only needed for establishing the local equilibrium in the normal region of the junction. If the coupling is very weak, this will influence the transient time $t=d/(\varepsilon v_F)$ to reestablish equilibrium if the bias voltage is changed, i.e. influence the switching speed.
The current gain in the JOINT can be defined as $G = I_r/I_1 \sim D/\varepsilon$ for the regular current (see also \cite{IlBag98}, and  $G = I_a/I_1  \sim \sqrt{DR}/\varepsilon$ for the anomalous nonequilibrium current \cite{WSSpatent98}. For $\varepsilon \ll 1$ this current gain can be become very large (recently demonstrated experimentally for the regular current \cite {SMNLHCHL98}).

To build a transistor utilizing the anomalous current, a Josephson interference transistor (JOINT) \cite{WSSpatent98},  is obviously going to be very demanding, and will require delicate control of the junction parameters. At least in principle, by properly gating the junction one might be able to tune the effective length of the junction, the scattering phase shifts and the coupling to the injection lead in order to optimize the junction to reveal the anomalous Josephson effect.

\section {Non-equilibrium population of Andreev levels via EM radiation.}

In optical physics it is common practice to control level populations by pumping with resonant electromagnetic (EM) radiation.

Similarly, the connection between the Andreev level energy $E(\phi,D,\beta)$ and the Josephson current $I(E)=(2e/\hbar)(dE/d\phi)$ should make it possible to control the Josephson current by using resonant EM radiation to pump between Andreev levels 
(Fig. 3), modulating the energy $E(\phi,D,\beta)$ and controlling the level populations $n(E)$. In principle one only has to modulate the phase difference $\phi$ or the normal transparency $D$ of the junction.

There are different mechanisms of coupling Andreev levels to external EM fields.
In general, a longitudinal electric field component oriented along the junction will create a time-dependent potential difference $V(t)$ across the junction and induce a time-dependent phase difference $\phi(t) = \phi + 2e \int V(t) dt $ by virtue of the Josephson relation. 
Another coupling mechanism, using e.g. a transverse electric field component, involves manipulation of the normal electron scattering properties of the junction, i.e. junction transparency $D$, electron density, etc. Such possibilites should exist in gate-controlled S-2DEG-S devices and mechanically controllable break junctions. 

The resulting level vibration will induce interlevel transitions, and the redistribution of excitations among the bound Andreev levels  will lead to a change of the Josephson current.
The matrix element for the interlevel transitions depends on the mechanism of the coupling and the junction geometry. In the simplest case $-$ voltage driven interlevel transitions in a short one-mode constriction with a single pair of Andreev levels (Fig. 3a) $-$ the matrix element has the form \cite{SWB93}
$\lambda=(\Delta/\omega)^2\sin^2(\varphi/2)D\sqrt{R}$. 
The matrix element for junctions with more complex geometry is given in \cite{Gor95}. It is important to mention that normal electron back scattering by the junction is essential for transitions between the Andreev levels: in purely ballistic constrictions ($R=0$) the interlevel transitions are forbidden due to conservation of the momentum of electrons at the Fermi level. 

The current response strongly depends on the way of switching on the radiation. Adiabatic turning on a flat-top EM pulse will suppress the Josephson current $I=\sum I(E)n(E)$ in a short constriction due to equalization of the level populations $n(E)$,
\begin{equation}
I/I_0= 1- \left[(\lambda eV)/ \hbar (\delta/2+\Omega)\right]^2
\end{equation}
where $\delta$ is the deviation from resonance (detuning), and $\Omega = \sqrt{(\lambda eV)^2 + \delta^2/4}$ is the Rabi frequency. Exactly at resonance, $\delta=0$, the current will be entirely suppressed. 

Since EM radiation induces a highly coherent mixture of Andreev states,
the off-diagonal part of the density matrix may give rise to oscillating contributions
to the current (Rabi oscillations). Therefore, if the EM power is turned on suddenly, the level populations $n(E)$ will oscillate between the lower level (the initial state) and upper level, causing Rabi oscillations in the Josephson current,
\begin{equation}
I(t)=I_0\left[1-2\left(\lambda eV/\hbar\Omega\right)^2\sin^2 \Omega t\right]
\end{equation}
The regime of current oscillation is sensitive to dephasing, which makes it possible to measure dephasing times.

Due to the selective character of the control, the lower bound for the power is only limited by the level width, originating e.g. from inelastic interaction with phonons. 
Since harmonic EM radiation only affects the resonant pair of Andreev states, the current control is most efficient if the number of Andreev states involved in the Josephson transport is small, i.e. short junction length and small number of normal transport electron modes.

In order to suppress the Josephson current, the members of the resonant pair of Andreev levels must carry current in opposite directions (Fig. 3a); otherwise nothing will happen. One may however ask whether it is possible to {\em enhance} the Josephson current by pumping with EM radiation. The answer is in principle yes: by using a suitably tuned SN-DBS-NS junction to establish an Andreev split-level scheme with $E\sim\sqrt{D}$ dispersion, as shown in Fig. 3b, one can kill the current from one of the levels by optical pumping, leaving an uncompensated "giant" Josephson current from the unpumped level. 

Morever, in a long junction filled with Andreev levels, it should be possible by resonant optical pumping to create a Josehpson current.
In junctions where $l_{\phi}>>\xi_{T}$, the Josephson current dies on a lengthscale much shorter than the phase breaking length of the electrons, due to cancellation among bound states and between the net bound state current and the continuum current.  In such a long junction with vanishing equilibrium Josephson current, by pumping between two Andreev levels, and therefore effectively killing the current from one bound state, it should be possible to unbalance the cancellation and create a mesurable Josephson current of one bound state arising from uncompensated continuum current (see also Ref. \cite{Argaman}).
\begin{figure}
\begin{centering}
\psfig{figure=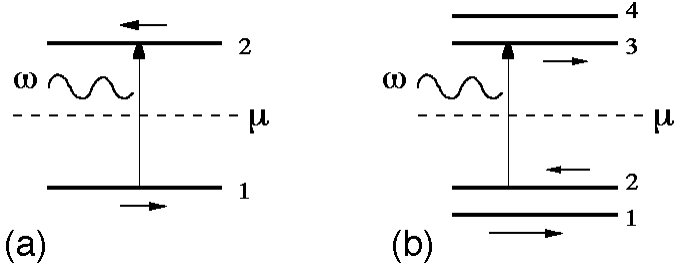,height=3.5cm}
\caption{(a) Quenching of the Josephson current via level saturation, equalizing the occupation numbers of leves 1 and 2. (b) Current enhancement by disclosing the nonequilibrium giant Josephson current of a single Andreev level (1): saturation of the $(2)\rightarrow(3)$ transition kills the opposing current of the other level (2) of the split pair.}
\end{centering}
\end{figure}

\section{Concluding remarks}

The recent progress in the field of superconducting mesoscopic junctions, with theory and experiment going hand in hand, has lead to the realization of elementary Josephson transistors which demonstrate the feasibility of controlling Josephson current. So far, only a few possibilities have been accomplished, mainly involving quenching of equilibrium Josephson current using electrostatric gates or current injection. Some interesting new possibilities involve {\em switching on} a Josephson current, using injection of normal current or electromagnetic radiation to create nonequlibrium conditions. These possibilities will require experimental control over multiterminal junctions with very few transport modes, and will present a major challenge for experimental investigations.\\\

Acknowledgement$-$This work has been supported by the Swedish grant agencies NFR, TFR, NUTEK, and the Japanese grant agency NEDO.


\begin{thebibliography}{999}
\bibitem{barrier}
In this paper we will in general use the term "barrier" to mean a flat-top potential hill in the band bottom, induced by intrinsic junction properties or external gates. If the top of this barrier exceeds the Fermi level we may refer to it as a "tunnel barrier" for the sake of clarity.
\end{thebibliography}
\end{document}